# ANISOTROPIC DEPENDENCE OF GIANT MAGNETO-IMPEDANCE OF AMORPHOUS FERROMAGNETIC RIBBON ON BIASING FIELD

B. KAVIRAJ[*]

**Magnetism Laboratory,** *Department of Physics and Meteorology, Indian Institute of Technology,*

*Kharagpur 721302, India*

*bhaskar@phy.iitkgp.ernet.in*

S.K. GHATAK

**Magnetism Laboratory,** *Department of Physics and Meteorology, Indian Institute of Technology,*

*Kharagpur 721302, India*

*skghatak@phy.iitkgp.ernet.in*

The magneto-impedance (MI) in amorphous ribbon of nominal composition $Fe_{73.5}Nb_3Cu_1Si_{13.5}B_9$ has been measured at 1MHz and at room temperature for different configurations of exciting a.c and biasing d.c. fields. A large drop in both resistance and reactance is observed as a function of d.c magnetic field. When the d.c and a.c fields are parallel but normal to the axis of ribbon, smaller magnetic field is needed to reduce the impedance to its small saturated value compared to the situation when fields are along the axis of ribbon. Larger d.c. field is required to lower the impedance when the d.c field acts perpendicular to the plane of the ribbon. Such anisotropy in magneto-impedance is related to the anisotropic response of the magnetization of ribbon. The large change of impedance is attributed to large variation of a.c permeability on the direction and magnitude of the dc biasing field.

Keywords: Magneto-impedance: Transverse permeability; Amorphous ferromagnet.

## 1. Introduction

The giant magneto-impedance (GMI) phenomenon refers to the observation of a large change in impedance of a soft magnetic material under the influence of d.c. magnetic field. The amorphous transition metal-metalloid ferromagnetic alloys are first systems where the GMI has been observed[1-5]. Although the effect has been found in amorphous alloys of different geometries (wire, ribbon, nanowire and film) the amorphous wire stands out among the GMI system due to magnetic softness and favorable geometry. The very large sensitivity of GMI on low magnetic field, stress and thermal treatment makes this effect very promising for technological applications [6-7].

It is generally understood that the combined effect of electromagnetic screening and magnetization dynamics in a ferromagnetic conductor leads to the GMI effect[8-10]. The magnetization dynamics depends upon the frequency and amplitude of excitation a.c field, biasing d.c. field, the magnetic anisotropy and their relative orientation. In amorphous state of transition metal-metalloid ferromagnet, the intrinsic magnetic anisotropy is predominantly magnetostrictive in character and the easy axis of magnetization thus depends on magnetostrictive coefficient and stress frozen during quenching process. Normally the easy axis of magnetization lies in the plane of ribbon and the magnetization dynamics differs when biasing field direction varies with respect to ribbon long axis. This then leads to anisotropic behaviour of GMI. In this communication, this anisotropic behaviour is reported in amorphous ribbon of nominal composition $Fe_{73.5}Nb_3Cu_1Si_{13.5}B_9$. A very sharp variation of GMI with d.c. field is observed when biasing and excitation fields are parallel to long axis of the ribbon and very broad variation results when d.c field is perpendicular to ribbon plane.

---

[*] Department of Physics and Meteorology
Indian Institute of Technology
Kharagpur 721302
India.

## 2. Experimental

The amorphous ribbon of nominal composition $Fe_{73.5}Nb_3Cu_1Si_{13.5}B_9$ was used. The real and imaginary components of impedance, $Z = R + jX$, where reactance $X = \omega L$ (L being the inductance of the sample) are measured by Q-meter (Model- Hewlett Packard) at a frequency of 1MHz. The sample is placed within a small secondary coil of rectangular geometry (100 turns) which carries the excitation current and length of the sample is larger than that of the secondary coil. The sample is then located within poles of electromagnet that produces the d.c. biasing magnetic field, $H_{dc}$. The sample is oriented such a way that its long dimension is normal to the Earth's magnetic field. The longer dimension of the sample is either perpendicular (hereafter referred as sample-1) or parallel (referred as sample-2) to the rolling direction of the ribbon. The lengths of the samples are 1.5 cm and width 3mm. The dimensions for both cases are kept same so that the demagnetizing effect remains identical. The exciting a.c. field 'h' is always parallel to the longer dimension of sample and its magnitude is around 100mOe. The response around this value of exciting field is found to be linear. The amplitude of the a.c field remains constant at all biasing fields. All the measurements are performed at room temperature. The d.c. biasing field is either parallel to or perpendicular to h, and in the latter case it is either in or out of the plane of sample and these arrangements are further referred as A,B and C respectively. In short, the alphabets 'A', 'B' and 'C' will relate to the orientation of d.c field $H_{dc}$ with respect to the a.c field h (along axis of ribbon). 'A' stands for the case when $H_{dc}$ and h are parallel and in the plane of ribbon. 'B' stands for the case when $H_{dc}$ and h are normal to each other but both in the plane of ribbon whereas 'C' represents the case when $H_{dc}$ and h are normal to each other and $H_{dc}$ being normal to the plane of the ribbon. The numeric '1' and '2' represents the case when the sample is cut normal to the ribbon axis and along the ribbon axis respectively. In this nomenclature, Sample-2A corresponds to the situation where the sample's longer dimension is along the axis of ribbon and h and $H_{dc}$ are also along the axis of the ribbon. The real and imaginary parts of the magnetic impedance are presented in terms of $\left(\frac{\Delta R}{R}\right)\% = \left((R(H_{dc}) - R(0))/R(0)\right) \times 100$ and $\left(\frac{\Delta X}{X}\right)\% = \left((X(H_{dc}) - X(0))/X(0)\right) \times 100$.

## 3. Results & Discussions

Fig.-1a shows the results of $(\Delta R/R)\%$ and $(\Delta X/X)\%$ of MI of the sample (sample –1A) cut normal to ribbon axis and the biasing dc field $H_{dc}$ and the excitation field h are parallel to each other and in a direction parallel to the length of the sample.

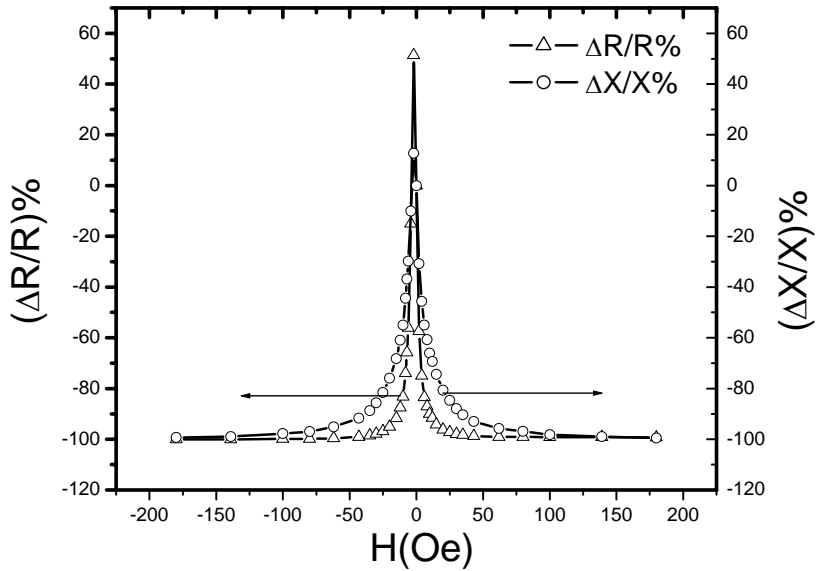

Fig 1(a)

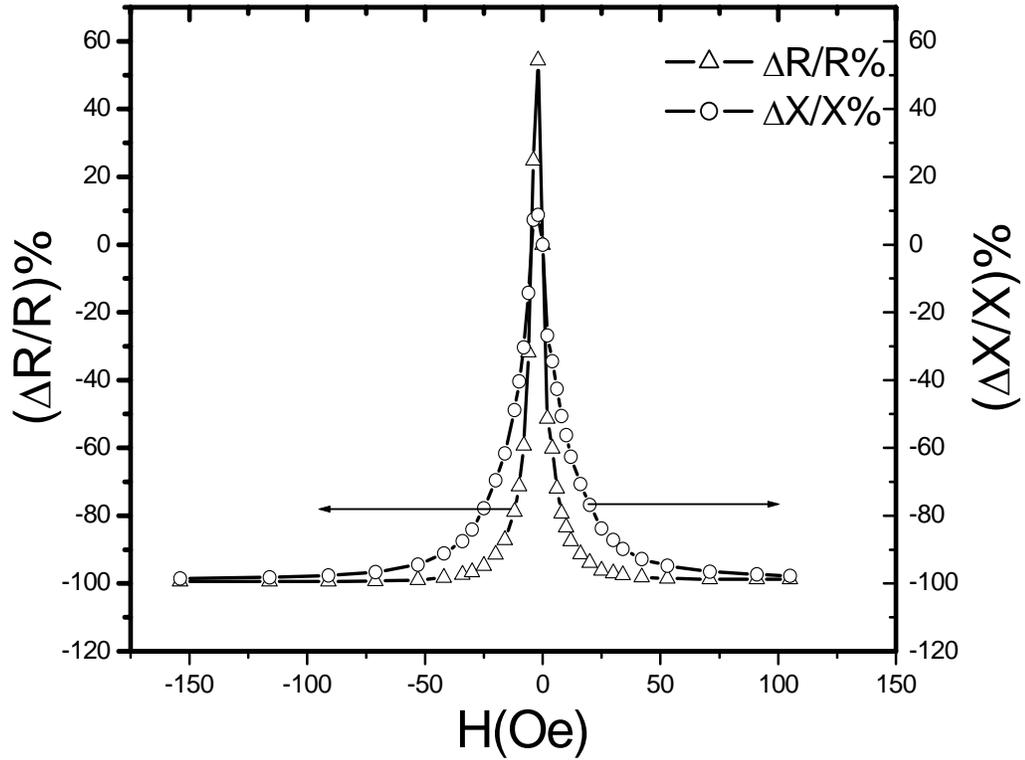

Fig 1: Percentage relative changes in resistance (ΔR/R) % and reactance (ΔX/X) % with $H_{dc}$ parallel to h. Fig (a) is for the sample cut normal to ribbon axis (sample 1A), and (b) is for the sample cut along the long axis (sample-2A).

It is seen that the real part of MI falls sharply with magnetic field and saturates to its very low value at small field (H~25Oe). The behaviour of the reactive component of impedance is similar at low field but saturates at a slightly higher field. At high field (around 100 Oe) the extent of fall in both (ΔR/R) % and (ΔX/X) % is nearly same and around 98%.

In Fig-1b the results for (ΔR/R)% and (ΔX/X)% are displayed for the sample (Sample-2A) with longer dimension along ribbon axis, and h and $H_{dc}$ are parallel to length of the sample. The field dependence of (ΔR/R)% and (ΔX/X)% are similar to the previous sample but the field needed to attain the saturation value of MI is higher in Sample-2A thus making the MI curves of Sample-2A less sharp than those of Sample-1A. In all cases the maximum appears at non-zero (but close to zero) field that depends on the configuration.

Fig2 (a and b) depict the field dependence of magneto-impedance when dc biasing field $H_{dc}$ is perpendicular to the exciting ac field (h), both being in the plane of the ribbon . In terms of nomenclature, Fig2a and 2b refer to Sample–1B and -2B respectively.

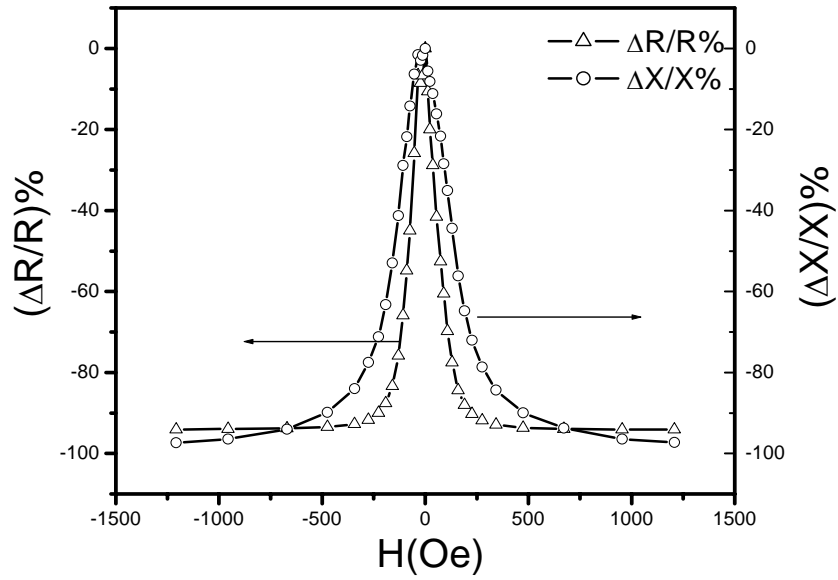

Fig 2(a)

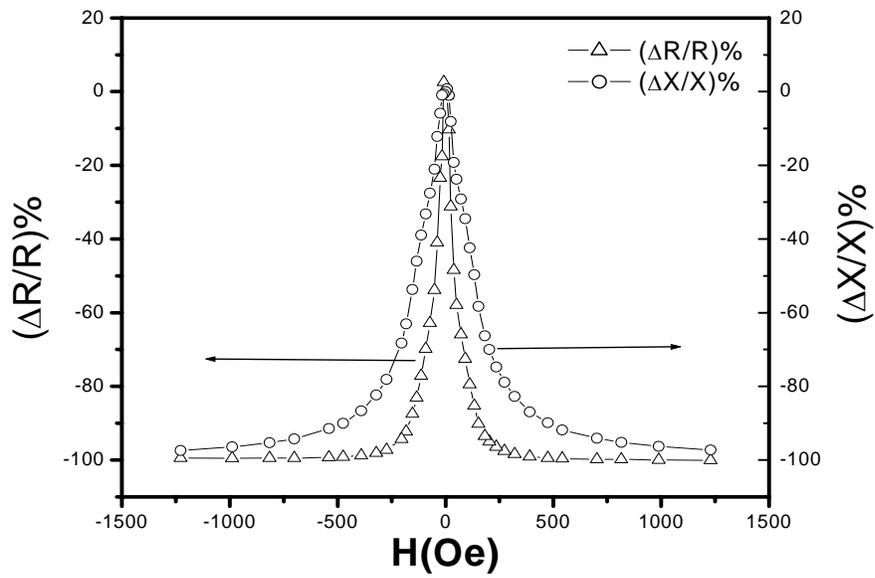

Fig2(b)

Fig 2: Percentage relative changes in resistance(ΔR/R)% and reactance(ΔX/X)% with $H_{dc}$ perpendicular to h and in the plane of ribbon. Fig (a) is for the sample cut normal to ribbon axis (sample 1B), and (b) for the sample cut along the long axis (sample-2B).

Both components of MI for the Sample –1B vary sharply for small $H_{dc}$, however differences appear at higher field. The variation of resistance almost vanishes beyond 500 Oe, and that of reactance persists over higher range of field. Much sharper variation is found for the Sample-2B and maximum change is found to be around 98% (Fig.-2b). Note that the fields that are required to saturate the impedance are higher in this configuration of fields compared to those for the situation $H_{dc} \parallel h$ ( Fig.-1a and 1b).

The results for MI when $H_{dc}$ is normal to the plane of ribbon are given in Fig.-3a and 3b designated as samples 1-C and 2-C respectively.

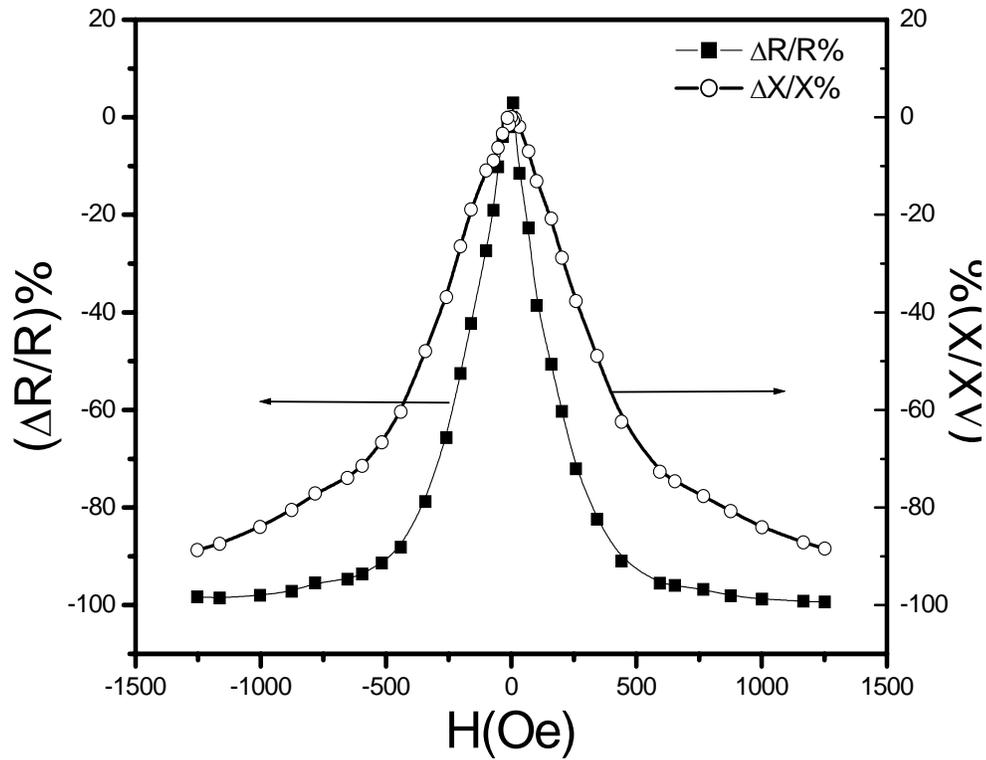

Fig 3(a)

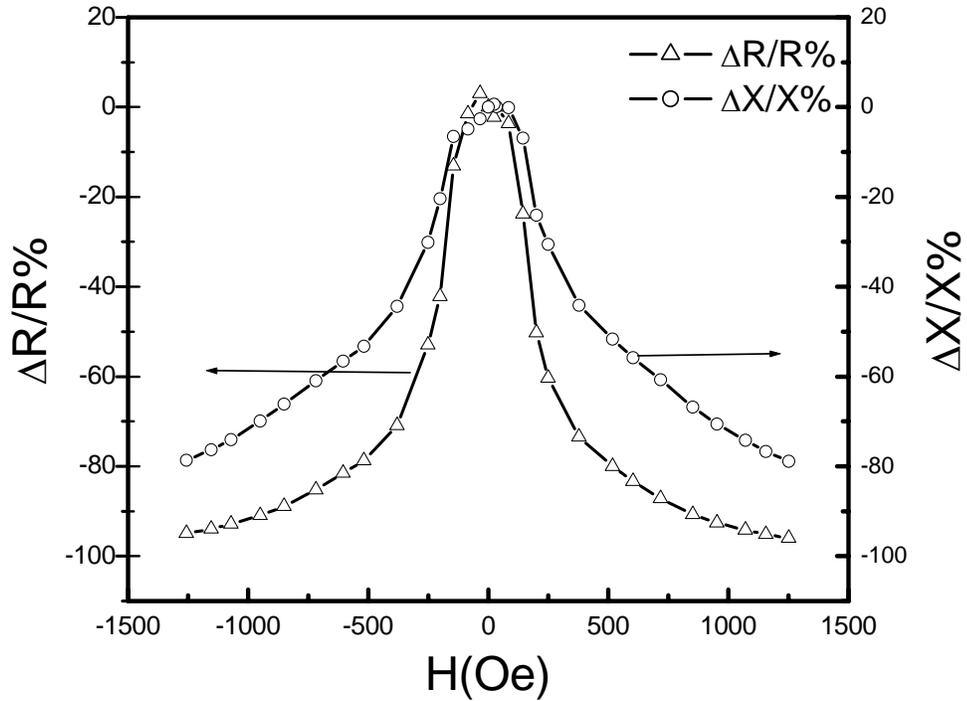

Fig 3: Percentage relative changes in resistance (ΔR/R)% and reactance (ΔX/X)% with $H_{dc}$ perpendicular to h and out of plane of ribbon. Fig (a) is for the sample cut normal to ribbon axis (sample 1C), and (b) is for the sample cut along the ribbon axis (sample-2C).

In this crossed condition of exciting and biasing fields both (ΔR/R)% and (ΔX/X)% vary slowly and do not saturate up to 1.25KOe. The difference between (ΔR/R)% and (ΔX/X)% for given field is larger compared to earlier situations. For sample 2C, both (ΔR/R)% and (ΔX/X)% vary little for small $H_{dc}$ (Fig.3b) and larger drop of (ΔR/R)% occurs around 150 Oe. The reactance shows much slower variation with field.

The above results are obtained by scanning the field from negative to positive direction. The hysterisis behaviour is found when the field is scanned in both directions. Fig 4(a) and 4(b) depicts the small hysteretic effect in the dc field response of the resistive and reactive parts of magneto-impedance of the amorphous ribbon when the field is scanned in forward (increasing) and reverse (decreasing) direction for the sample- 2A.

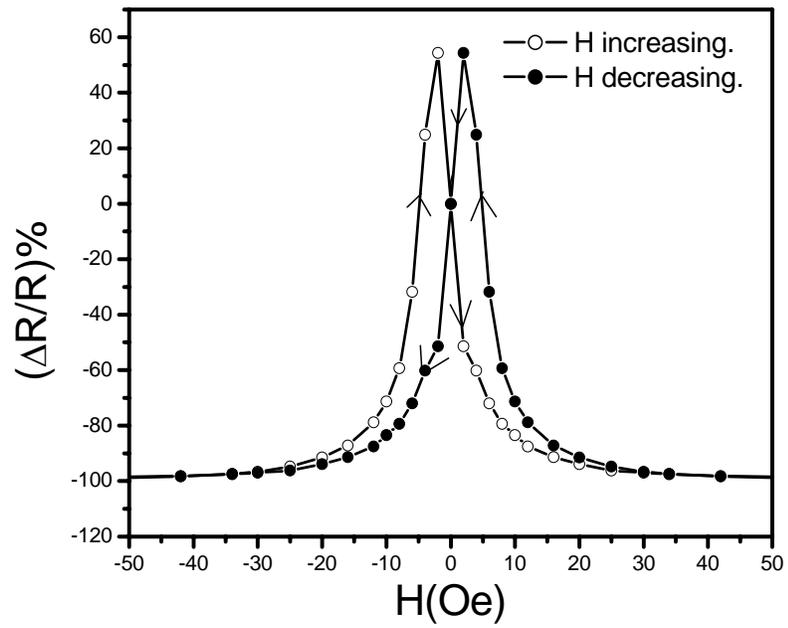

Fig 4(a)

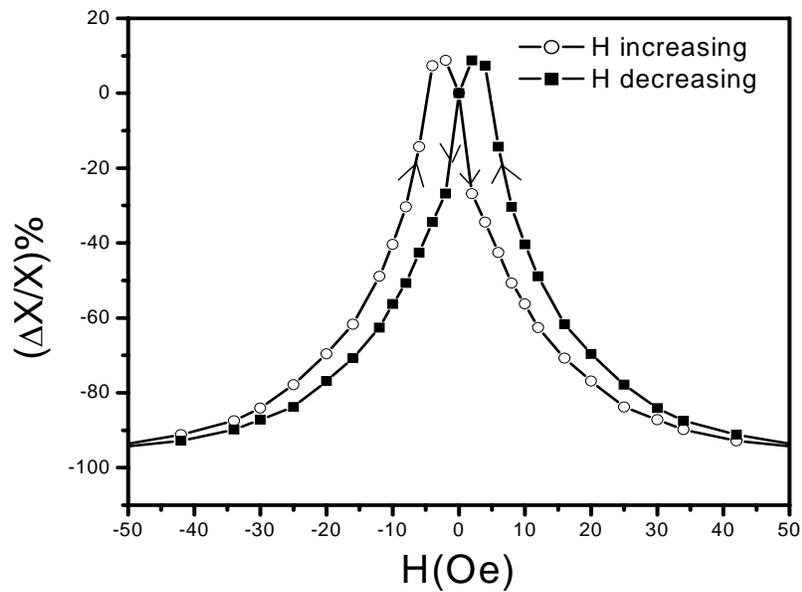

Fig 4(b)

Fig.4 .:Hysterisis behaviour of resistance (a) and reactance (b) for the sample 2A For increasing direction of the field, the peak of resistance (reactance) appears at -2.4 Oe and for decreasing case it occurs at 2.4 Oe . In all the cases similar hystersis behaviours have been observed.

It is evident that the MI response of the sample depends on the relative orientation of exciting and biasing fields. In order to examine in detail the orientational dependency of MI, the measurements are carried out by varying angle ($\theta_0$) between h and $H_{dc}$ . Fig 5 (a) and (b) depicts the dc field dependence of R (real) and X (imaginary) respectively for the sample –2 (longer dimension parallel to the rolling direction) at various values of $\theta_0$ . Here $\theta_0 = 0^0$ corresponds to the situation where the fields are parallel to ribbon axis, and $\theta_0 = 90^0$ corresponds to the case where $H_{dc}$ is perpendicular to ribbon axis (and to h) .

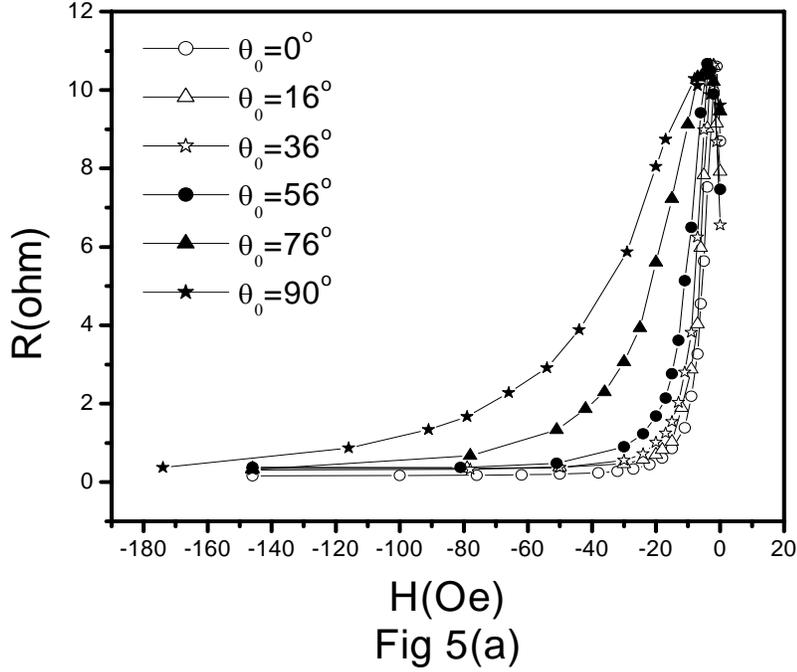

Fig 5(a)

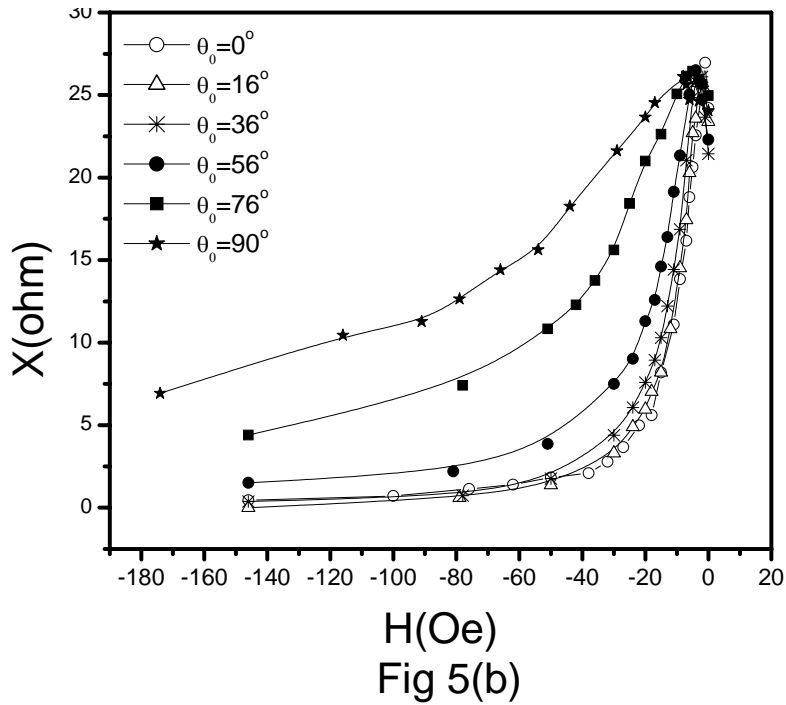

Fig 5(b)

Fig. 5: Variation of real (a) and imaginary (b) parts of impedance with angle $\theta_0$ between $H_{dc}$ and h for sample cut parallel to ribbon axis (sample-2).

The field $H_{Zmax}$ where R and X exhibits peak decreases from 8Oe to 1Oe as $\theta_0$ goes from $90^0$ to $0^0$. The variation $H_{Zmax}$ with $\theta_0$ is non-linear and is shown in Fig.6.

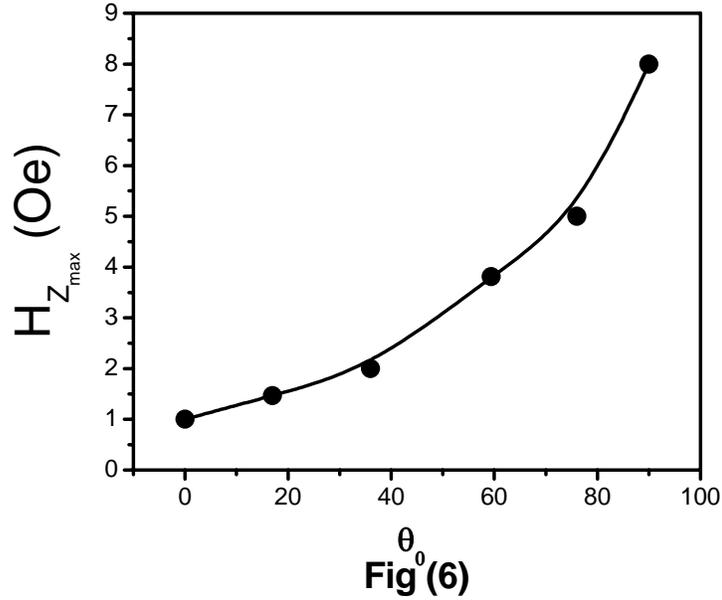

Fig.6 : Angular variation of $(H)_{Z\,max}$ for sample-2.

At low fields, steep increase of MI with field is nearly same for all orientation. However, the fall of the MI beyond $H_{Zmax}$ depends on $\theta_0$ and for a given $H_{dc}$, MI increases with $\theta_0$. The field needed for saturation is also higher at higher $\theta_0$ as evident from Fig (5).

The general trend of results can be understood from the fact that MI depends upon the a.c. response of the sample in presence of a bias field. From the Maxwell's equations of electromagnetic fields in a conducting media, the electric $e$ and magnetic $h$ fields can be expressed as [8-11]

$$\vec{\nabla} \times \vec{h} = \sigma \vec{e} \qquad (1)$$

$$\vec{\nabla} \times \vec{e} = -i\omega\mu \vec{h} \qquad (2)$$

Here 'ω' is the angular frequency of the excitation current, 'μ' is the permeability tensor and 'σ' is the conductivity of the medium. Considering solutions of (1) and (2) in linear approximation, the impedance can be expressed as $Z(\omega, H_{d.c}) = V/I$, where voltage V develops across the sample coil that carries a.c. current $I = I_0 \exp(-j\omega t)$.

The impedance Z for a ribbon with thickness d << length and width of ribbon can be expressed in terms of complex wave vector k as

$$Z = -j\omega X_0 \mu_e \frac{\tanh(kd)}{kd} \quad (3)$$

where $X_0$ is the reactance of empty sample coil, $\mu_e = \mu_e' + j\mu_e''$ is the effective permeability of the material. The current flowing through the coil generates the axial magnetic field $\mathbf{h_z}$ and induction $\mathbf{b_z} = \mu_e \mathbf{h_z}$. The wave vector $k$ is given as,

$$k = \left(\frac{1+j}{\delta_m}\right) \qquad \delta_m = \left[\frac{2}{\omega \mu_0 \mu_e \sigma}\right]^{1/2} \quad (4)$$

where $\delta_m$ is skin depth, $\mu_0 = 4\pi \times 10^{-7}$ H/m and $\sigma$ = d.c conductivity of the material. It follows from Eq.(3) and Eq.(4) that the magneto-impedance of the material is determined by the magnetic response $\mu_e$. The permeability depends upon the equilibrium orientation of magnetization which is governed by the anisotropy and external fields. In absence of d.c field, the response to a.c. excitation is very large for soft ferromagnetic system (low magnetic anisotropy) and this means large permeability. In the limit $\delta_m << d$, (3) can be approximated as

$$Z \approx -(j-1) X_0 \mu_e \left(\delta_m / d\right) \quad (5)$$

So the impedance varies as $\mu_e^{1/2}$ and is large for material with large $\mu_e$. The other limit can be achieved in presence of large biasing field ($H_{dc} >> h$) that reorients magnetization along $H_{dc}$ and the a.c. magnetization parallel to h is drastically reduced. This in turn increases $\delta_m$ and $Z \approx -jX_0 \mu_e$. Here $\mu_e$ is the permeability in presence of large biasing field and is reduced by an order of magnitude compared to that at $H_{dc} = 0$. The large negative MI is therefore the result of additional screening of e.m field in a magnetic metal. The additional screening current depends upon the magnitude and rate of variation of a.c. magnetization and hence MI increases with frequency of exciting field. The magnetization induced by a.c field depends upon the equilibrium configuration of magnetization, which in turn is determined by relative orientation of anisotropy and bias fields. It is assumed that the magnetization vector lies in the plane of the ribbon and the easy axis of magnetization marked by anisotropy field $H_a$ is not along the axis of the ribbon (Fig. 7).

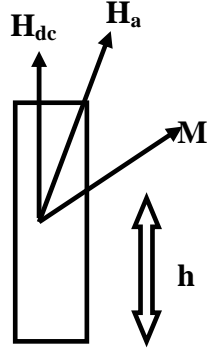 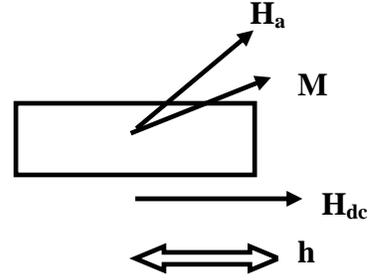

**Fig 7(a)**  **Fig 7(b)**

Fig.7: Diagram showing relative orientation of magnetization M and external fields $H_{dc}$, h and anisotropy field $H_A$. The long direction of the sample is normal (Fig. 7a) and parallel (Fig. 7b) to ribbon axis respectively.

When the sample's longer dimension is normal to ribbon axis and fields are parallel (sample -1A) the torque due to $H_{dc}$ on magnetization M adds to that due to $H_A$ and this results faster alignment of M along $H_{dc}$. So as $H_{dc}$ increases, M approaches to fully aligned state and in the limit h<<$H_{dc}$, the induced a.c magnetization becomes very small. Therefore, the permeability, which is a measure of the response of sample to a.c. field, decreases rapidly with increase in $H_{dc}$. This in turn increases the skin depth resulting in a rapid drop of MI (Fig.1a). For the sample whose longer dimension is parallel to ribbon axis along which both fields are applied, the torque due to $H_A$ opposes that due to $H_{dc}$. As a result alignment of the magnetization changes slowly and the permeability also decreases slowly with increase in bias field. This result in a slower variation of MI (Fig.1b) compared to the earlier situation. In crossed field configuration –B, the faster variation of MI is expected in sample-2 compared to that for sample-1.This is due to the fact that the torques due to $H_{dc}$ and $H_A$ on M are co-linear in former case and are anti-parallel in sample-1. Since h is normal to $H_{dc}$ the induced magnetization is higher compared to the parallel configuration. This leads to higher values of permeability for a given $H_{dc}$ and thus slower variation of MI in crossed field configuration. This explains qualitatively higher width of MI variation in B compared to that in A. In crossed configuration –C, very large field is needed to orient M in out-of plane of ribbon. In this case shape anisotropy field is very dominant due to high value of demagnetization factor and much higher biasing field is required to saturate the MI as has been observed (Fig.3).

## 4. Conclusions

In this communication, the magneto-impedance of ferromagnetic amorphous ribbon with nominal composition $Fe_{73.5}Nb_3Cu_1Si_{13.5}B_9$ has been measured as a function of applied dc magnetic field for different configurations of exciting ac and biasing dc fields. A giant magneto-impedance has been observed at low biasing field. The results clearly demonstrate that the extent of decrease in impedance depends upon the relative orientation of dc and anisotropy fields. It also shows that the easy axis of magnetization in this ribbon is nearly perpendicular to ribbon axis. As the sample has small positive magnetostriction the results suggest the presence of compressive stress in as-quenched ribbon. The measurement of GMI can therefore be utilized to find easy axis of magnetization and possible origin.